\documentclass[fleqn,10pt]{wlscirep}

\title{Transportation dynamics on coupled networks with limited bandwidth}

\author[1,*]{Ming Li}
\author[1]{Mao-Bin Hu}
\author[2,$\dag$]{Bing-Hong Wang}
\affil[1]{School of Engineering Science, University of Science and Technology of China, Hefei, 230026, P. R. China}
\affil[2]{Department of Modern Physics, University of Science and Technology of China, Hefei, 230026, P. R. China}

\affil[*]{minglichn@ustc.edu.cn}
\affil[$\dag$]{bhwang@ustc.edu.cn}

\begin{abstract}
The communication networks in real world often couple with each other to save costs, which results in any network does not have a stand-alone function and efficiency. To investigate this, in this paper we propose a transportation model on two coupled networks with bandwidth sharing. We find that the free-flow state and the congestion state can coexist in the two coupled networks, and the free-flow path and congestion path can coexist in each network. Considering three bandwidth-sharing mechanisms, random, assortative and disassortative couplings, we also find that the transportation capacity of the network only depends on the coupling mechanism, and the fraction of coupled links only affects the performance of the system in the congestion state, such as the traveling time. In addition, with assortative coupling, the transportation capacity of the system will decrease significantly. However, the disassortative coupling has little influence on the transportation capacity of the system, which provides a good strategy to save bandwidth. Furthermore, a theoretical method is developed to obtain the bandwidth usage of each link, based on which we can obtain the congestion transition point exactly.
\end{abstract}

\begin{document}

\flushbottom
\maketitle
\thispagestyle{empty}

\section*{Introduction}

To meet the needs of communication development, more and more communication networks have been built over the last few years. However, in reality, due to the exorbitant cost of laying special lines and their energy supplies, a new communication network often attaches to some power networks, or partially couples with a existing communication network. Obviously, this construction mechanism makes the new network does not have a stand-alone function and efficiency. For the former case, a cascading failure caused by the interdependence between the power network and the communication network could destroy the whole system more easily than that without coupling\cite{buldyrev2010catastrophic}. The so-called interdependent networks are the popular model to perform such study\cite{Boccaletti20141,Kivela01092014}. For the latter case, on one hand, the coupling of two or more communication networks enables the message, opinion, or other information to spread from one network to another, which could facilitate the spreading and the emergence of cooperation \cite{PhysRevLett.110.028701,0295-5075-97-4-48001,gomez2012evolution,jiang2013spreading,PhysRevE.86.026106,PhysRevE.88.022801}. On the other hand, sharing some physical devices could reduce the transportation efficiency of the communication networks. For example, in reality, some communication companies often share some communication lines to save costs, although the nodes in these networks may be totally different. Within the process of communication, the common lines will share a bandwidth. When the common bandwidth has been used by one network, the other networks can not use it at the same time. Uncovering the effects of this coupling on transportation dynamic will be helpful for real-world communication network design and bandwidth allocation, which is the focus of this paper.

A basic model of transportation dynamic on networks is the one proposed by Ohira and Sawatari\cite{PhysRevE.58.193}. In this model, at each time step, information packets are generated with destination addresses, and transferred from one node to another toward their destinations following the given routings. With the increasing of the packet generation rate, this model can exhibit a phase transition from the free-flow state to the congestion state\cite{PhysRevLett.86.3196}. Based on this model, the dynamical properties of the communication system have been studied widely\cite{PhysRevLett.87.278701,PhysRevE.66.026704,PhysRevE.71.026125,PhysRevE.74.046106}, and a lot of routing strategies have been proposed to improve the transportation efficiency of a given network\cite{wang2007traffic}, such as increasing the transportation capacity and decreasing the traveling time\cite{PhysRevE.73.046108,PhysRevE.73.026111,PhysRevE.81.016113}. Some models and routing strategies have also been studied in the so-called interconnected networks\cite{0295-5075-102-2-28002,PhysRevE.89.062813,PhysRevLett.116.108701}. However, the interconnected networks in these models can all be understood as a large network with community structure, the sharing of physical devices between different networks has not been taken into account.

To appreciate what the sharing of physical devices in communication networks will mean, we will consider two transportation processes display on two networks simultaneously with some link coupling. The transportation in the two networks is limited by the link bandwidth. That is each link can only deliver a limited number of information packets at each time step\cite{PhysRevE.80.066116,0295-5075-79-1-14003,Ling20104825,PhysRevE.84.026116}. For two coupled links, they share a common bandwidth limitation, which just represents the sharing of physical devices between the two communication networks. We will show that the transportation capacity of the two networks could be different, i.e., the free-flow state and the congestion state can exist in the two networks, respectively. By the analysis of different coupling mechanisms, we find that the congestion transition point is dependent on the coupling mechanism, and independent of the fraction of coupled links. Based on that, we demonstrate that the disassortative coupling is a good strategy for saving bandwidth, for which we can use the bandwidth of one network to support the transportation of the two networks without reducing the transportation capacity of both networks. All these results are established through analytic theory with numerical support.

\section*{Results} \label{result}

\subsection*{Model} \label{model}

The transportation system is formed by two networks $A$ and $B$ with the same size $N$ and degree distribution. At each time step, each node generates a new information packet with probability $r$, i.e., $rN$ new packets in each network. The newly generated packet will be placed at the end of the queue of its origin node with a randomly chosen destination. At each time step, each node will deliver the packets in its queue toward their destinations, if the corresponding links have free bandwidth. Here, the bandwidth of a link is the maximum number of packets that can be delivered through the link at each time step. Without loss of generality, we set the bandwidth to be $1$ for all links, i.e., only one packet can be delivered at each time step. If the corresponding bandwidth limit is reached, the packet has to stay in the queue. The transportation of a node will be over, when none of its links have free bandwidth. For simplicity, all the packets are delivered in the networks following the shortest paths from the origins to the destinations. When a packet reaches its destination, it will be removed from the system.

\begin{figure}[ht]
\centering
\includegraphics[width=0.4\linewidth]{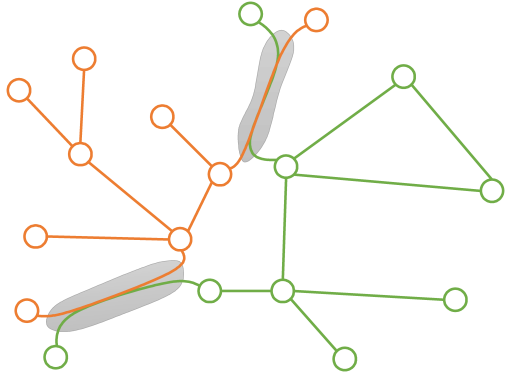}\caption{A schematic of two communication networks with link coupling. The two networks are presented by orange and green, respectively. The pairs of links indicated by the grey areas are the coupled links of the two networks. In the transportation process, two coupled links will share a common bandwidth, which can be used by both of the two networks but with a priority.}
\label{f1}
\end{figure}

To represent the coupling of the two networks, we assume that a fraction $\beta$ of the links of network $A$ are coupled with the links of network $B$ one by one (see Fig.\ref{f1}). In the transportation process, two coupled links are considered as one link, i.e., sharing bandwidth. To present the priority of the transportation process, we assume that network $A$ uses the common bandwidth first at each time step, and the remaining bandwidth (if any) can be used by network $B$.

In this paper we will study three coupling mechanisms, random, assortative and disassortative couplings. For random coupling, all the coupled links are chosen randomly from the two networks. For assortative (disassortative) coupling, the link with the largest betweenness of network $A$ is coupled with the link with the largest (smallest) betweenness of network $B$, the link with the second largest betweenness of network $A$ is coupled with the link with the second largest (smallest) betweenness of network $B$, and so on.

\subsection*{Measures}

To describe the transition from the free-flow state to the congestion state, we use the following order parameter
\begin{equation}
\eta=\lim_{t\rightarrow \infty} \frac{\Delta S}{rN\Delta t}=\lim_{t\rightarrow \infty} \frac{\langle S(t+\Delta t)-S(t)\rangle}{rN\Delta t}. \label{eta}
\end{equation}
Here, $S(t)$ is the total number of packets in the network at time step $t$, and $\langle \cdot\rangle$ means the average for different $t$ but the same $\Delta t$. It is easy to know that the maximum value of $\Delta S/\Delta t$ is $rN$, which gives the maximum $\eta=1$. Below the critical point $r_c$, $\Delta S/\Delta t=0$, i.e., at each time step, the number of packets arriving at their destinations is equal to $rN$.

As long as one link can not deliver all the packets that need to be delivered, the order parameter $\eta$ will be larger than zero and the system will turn into the congestion state. Therefore, to study this congestion transition, we also need to measure the transportation state of each link quantitatively. Here, we use the bandwidth usage $\mu_{ij}$ to represent the traffic load of link $i\rightarrow j$, which defines as the average number of packets passing through link $i\rightarrow j$ at each time step. Mathematically, it can be expressed as
\begin{equation}
\mu_{ij}=\frac{\sum_{t=t_1}^{t_2} S_{ij}(t)}{t_2-t_1+1},
\end{equation}
where $S_{ij}(t)$ is the number of packets passing through link $i\rightarrow j$ at the time step $t$. Since $S_{ij}(t)$ can only take value $1$ or $0$ in our model, the range of the bandwidth usage $\mu_{ij}$ is from $0$ to $1$. It's clear that the larger the bandwidth usage is, the heavier the load is. When $\mu_{ij}\rightarrow 1$, the link would be congested.

\subsection*{Transportation on single network}

\begin{figure}[ht]
\centering
\includegraphics[width=0.95\linewidth]{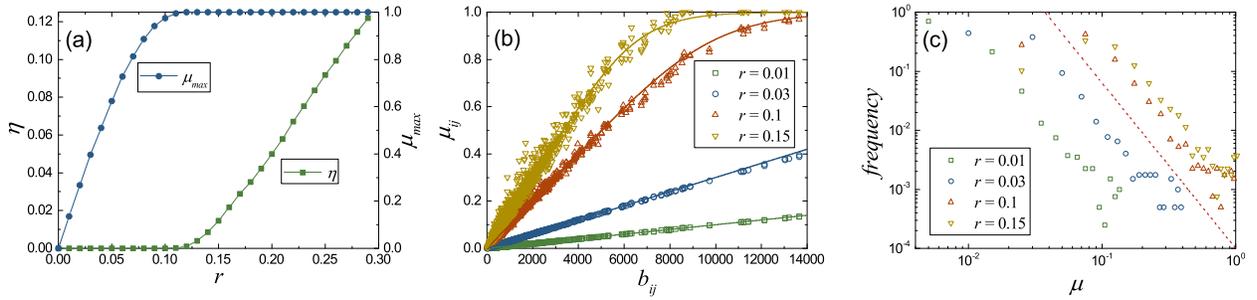}
\caption{The simulation results for transportation on single networks ($\beta=0$). In the simulation, the networks we used are BA networks with average degree $\langle k\rangle=4$ and size $N=1000$. (a) The order parameter $\eta$ as a function of the packet generation rate $r$. $\mu_{max}$ is the bandwidth usage of the link with the largest betweenness. (b) The bandwidth usages of links $\mu_{ij}$ are plotted as a function of their betweennesses $b_{ij}$. The corresponding lines are obtained by eqs.(\ref{mu}) and (\ref{ln3}). For small packet generation rate $r$, the bandwidth usage of link increases approximate linearly with the betweenness, which also satisfies eq.(\ref{mus}) (see the cases $r=0.01$ and $0.03$). (c) The frequency counts of the bandwidth usages $\mu$ shown in (b). The slop of the dash line is $-2.8$.}
\label{f2}
\end{figure}

For the convenience of analysis and discussion, we study the case $\beta=0$ first. In this case, there is no interaction between the two networks, so we just consider one of them. To model the topological properties of real communication networks, we use Barab\'{a}si-Albert (BA) network in the simulation\cite{Barabasi509}. As shown in Fig.\ref{f2} (a), the order parameter $\eta$ will take a non-zero value when the packet generation rate $r$ exceeds a critical value $r_c$, which is often called the transportation capacity of the system.

To get a better understanding of this congestion transition, we sort the bandwidth usages $\mu_{ij}$ of all the links by their betweennesses $b_{ij}$ in Fig.\ref{f2} (b). Together with the frequency counts of these bandwidth usages shown in Fig.\ref{f2} (c), one can find that when the bandwidth usages of most of links are very small, the bandwidth usage of the link with the largest betweenness $\mu_{max}$ will close to the maximum value $1$. From eq.(\ref{eta}), we can also find that only one link with $\mu\rightarrow 1$ will make the system go into the congestion state. Therefore, the transition point $r_c$ of the system is just the point above which $\mu_{max}=1$ (see Fig.\ref{f2} (a)). From this point of view, we can obtain the theoretical transition point for Fig.\ref{f2}, $r_c=0.112$ (see Sec. Method for details), which is in agreement with simulation results well.

\subsection*{Transportation on coupled networks}

\begin{figure}[ht]
\centering
\includegraphics[width=0.95\linewidth]{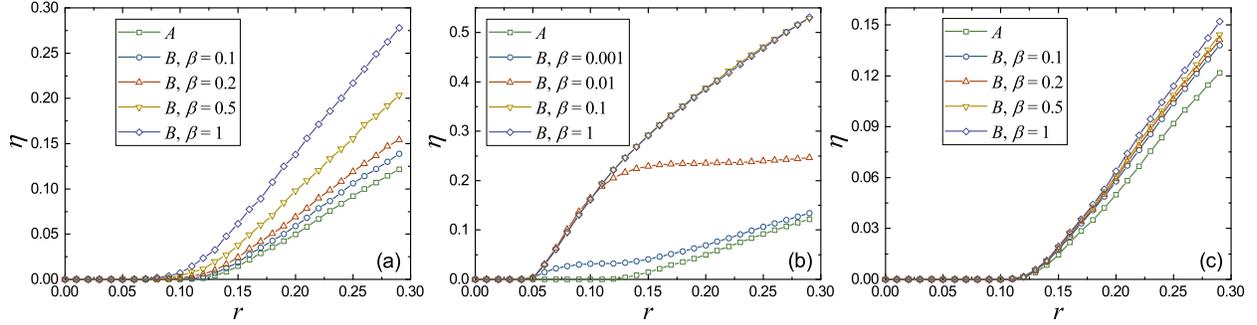}
\caption{The order parameter $\eta$ as a function of the packet generation rate $r$ for different coupling fractions $\beta$. In the simulation, both networks $A$ and $B$ are BA networks with the same average degree $\langle k\rangle=4$ and size $N=1000$. (a) Random coupling. (b) Assortative coupling. (c) Disassortative coupling.}
\label{f3}
\end{figure}

Since network $A$ has the priority in using the common bandwidth, the coupling will not affect its communication performance. Thus, we focus now on the performance of network $B$ in different coupling mechanisms. In Fig.\ref{f3}, we give the order parameter $\eta$ as a function of the packet generation rate $r$ for different coupling fractions $\beta$. We can find that for assortative coupling, a very weak coupling could reduce the transportation capacity $r_c$ of network $B$ significantly. However, the random and disassortative coupling, especially disassortative coupling, do not have serious effect on the transportation capacity $r_c$ of network $B$. This indicates that if the traffic of network $B$ is not so heavy, we need not to build links just for network $B$ and all its communication task can be attached to network $A$. This provides a good strategy to save costs for building new communication network.

\begin{figure}[ht]
\centering
\includegraphics[width=0.8\linewidth]{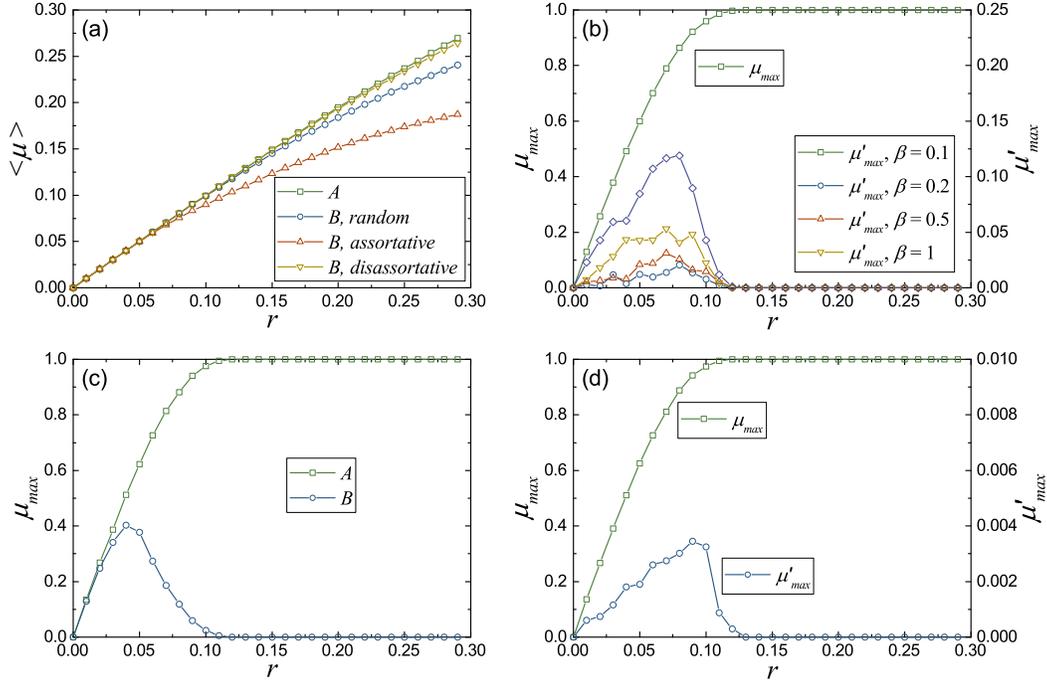}
\caption{The bandwidth usage of links for different coupling mechanisms. (a) The average bandwidth usage $\langle\mu\rangle$ for different coupled mechanisms with $\beta=1$. (b) Random coupling. The maximum bandwidth usage $\mu_{max}$ of the link with the largest betweenness in network $A$, and the bandwidth usage $\mu_{max}^\prime$ of its coupled link in network $B$ as a function of the packet generation rate $r$. In this case, the link with the largest betweenness is not always chosen as the coupled link. However, since this link always gives the smallest congestion transition point $r_c$, for the average of many realizations, we can treat the results as that the link with the largest betweenness is always the coupled link. (c) Assortative coupling. The maximum bandwidth usages $\mu_{max}$ of the links with the largest betweennesses in networks $A$ and $B$, which are coupled with each other, as a function of the packet generation rate $r$. (d) Disassortative coupling. The maximum bandwidth usage $\mu_{max}$ of the link with the largest betweenness in network $A$, and the bandwidth usage $\mu_{max}^\prime$ of its coupled link in network $B$, whose betweenness is the smallest one in network $B$, as a function of the packet generation rate $r$. In the simulation, both networks $A$ and $B$ are BA networks with the same average degree $\langle k\rangle=4$ and size $N=1000$.}
\label{f4}
\end{figure}

To figure out the capacity of network $B$, we also take account of the bandwidth usage as that of single networks shown in Fig.\ref{f4}. For random and assortative couplings, we can find that the average bandwidth usage $\langle\mu\rangle$ of network $B$ will deviate from that of network $A$ with the increasing of the packet generation rate $r$ (see Fig.\ref{f4} (a)). However, the average bandwidth usage for the disassortative coupling almost coincides with that of network $A$. This indicates that the disassortative coupling does not add too much restriction to network $B$ in using the common bandwidth, and makes the bandwidth allocation in the coupled system more reasonable. This explains why the system with the disassortative coupling shown in Fig.\ref{f3} has the largest transportation capacity.

Since the two networks have the same structure and traffic load, the coupling will not affect the transportation in network $B$, before the average bandwidth usage of network $A$ exceeds $0.5$. However, the simulation results shown in Fig.\ref{f4} (a) contradict this. Addressing this issue, we consider the heterogeneity of the bandwidth usage as that of single network. For network $B$, the bandwidth of a link is restricted by the coupling, and only the remaining bandwidth from network $A$ can be used. It is clear that the link with the largest bandwidth usage $\mu_{max}$ in network $A$ has the smallest remaining bandwidth. For convenience, we denote the bandwidth usage of this link in network $B$ as $\mu_{max}^\prime$.

In Fig.\ref{f4}, we give the bandwidth usages $\mu_{max}$ and $\mu_{max}^\prime$ as a function of the packet generation rate $r$. We can find that the bandwidth usage $\mu_{max}^\prime$ will decrease with the increasing of $r$, when the packet generation rate $r$ exceeds a value, at which $\mu_{max}+\mu_{max}^\prime\rightarrow1$. In this way, the critical value $r_c$ obtained by $\mu_{max}+\mu_{max}^\prime = 1$ just is the congestion transition point of network $B$, since it is smaller than that obtained by the link with the largest betweenness in network $B$. For the simulation results shown in Figs.\ref{f3} and \ref{f4}, $\mu_{max}^\prime$ only depends on the coupling mechanism. This explains that the congestion transition point $r_c^B$ in Fig.\ref{f3} takes the same value for different $\beta$. Furthermore, using $\mu_{max}+\mu_{max}^\prime= 1$, we can also obtain the transition point $r_c^B$ for the random, assortative and disassortative couplings, that is $0.078$, $0.037$ and $0.109$ for the simulation results shown in Fig.\ref{f3} (see Sec. Method for details).

\subsection*{In the congestion state}

All the above results suggest as long as only one link is congested, the system will go into the congestion state. This means that in the congestion state not all the packets will be stuck, and the ones with a routing containing no congested links can be delivered as usual. In other words, the free-flow and congested paths can coexist in the system. This is a quite common phenomenon in real communication networks.

For assortative coupling, all the links with larger bandwidths are chosen as the coupled links, so with the increasing of $r$, the coupled links are always congested firstly. In this case, we can classify the congested and free-flow links simply by whether or not they are coupled links. Therefore, we take the assortative coupling as an example to show the coexistence of the free-flow paths (containing no coupled links) and the congested paths (containing coupled links) in the system.

\begin{figure}[ht]
\centering
\includegraphics[width=0.6\linewidth]{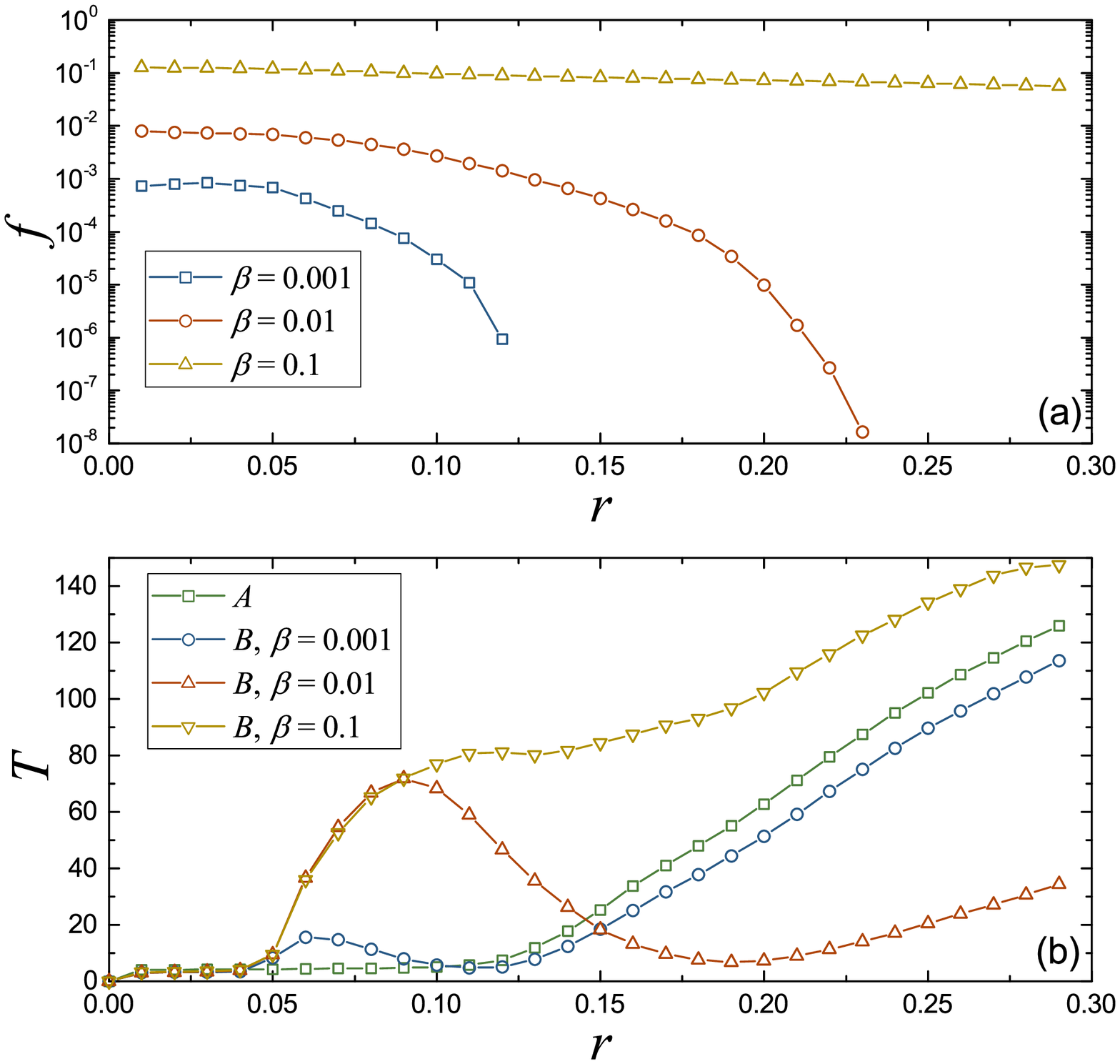}
\caption{The simulation results for showing the coexistence of the free-flow paths and the congested paths in the congestion state. In the simulation, both networks $A$ and $B$ are BA networks with the same average degree $\langle k\rangle=4$ and size $N=1000$. (a) The fraction $f$ of the packets that reach their destinations by passing through one or more coupled links in network $B$ for assortative coupling. (b) The average traveling time $T$ as a function of the packet generation rate $r$ for assortative coupling.}
\label{f5}
\end{figure}

In Fig.\ref{f5} (a), we give the fraction $f$ of the packets that reach their destinations by passing through one or more coupled links in network $B$ at each time step for assortative coupling. If the congestion happens in each link simultaneously, $f$ will be a constant depending on the coupling fraction $\beta$ and the link betweenness, regardless of the system is in the free-flow state or the congestion state. However, the results of Fig.\ref{f5} (a) show that when the packet generation rate $r$ exceeds $r_c^B$, $f$ decreases with the increasing of the packet generation rate $r$. This indicates that the free-flow and congested paths coexist in the congestion state.

Furthermore, in the congestion state, all the packets passing through some coupled links in network $B$ will wait a long time for being delivered, or never reach their destinations ($r>r_c^A$). Since only the packets that can reach their destinations contribute to the statistics of the average traveling time $T$, this creates an illusion that the packets will take less time to reach their destinations, when the packet generation rate $r$ closes to $r_c^A$ (see Fig.\ref{f5} (b)). As above analysis, this does not mean the traffic condition is improved with the increasing of the packet generation rate $r$, but the coexistence of the free-flow paths and the congested paths in the congestion state. For the other two coupling mechanisms, this coexistence can also be found in the congestion state. However, not all the congested links are the coupled links, so one needs to make a distinction between the free-flow and congestion links by other parameters, such as bandwidth usage. The results are similar, so we do not go into this in detail.

\section*{Conclusions}

In this paper we have proposed a transportation model to study the bandwidth sharing between different communication networks. Two coexistence phenomena have been found in this simple transportation model, both of which reflect the common problems in daily communication. One is the coexistence of the free-flow state and the congestion state in the two networks, the other one is the coexistence of the free-flow path and the congestion path in each network. According to our analysis, the former is caused by the priority of the two networks in using the common bandwidth, the latter is due to the heterogeneity of the traffic load on each link.

We also develop a theoretical method to obtain the congestion transition points of the two networks, which are in agreement with the simulation results very well. Both the theory and the simulation results indicate that the transportation capacity of the system depends on the coupling mechanism of the system, no matter how many number of links are coupled with each other. Furthermore, we find that the disassortative coupling has little influence on the transportation capacity of the system, which provides a good strategy to save bandwidth. We also point out that the coexistence of the free-flow path and the congested path in the congestion state could create the illusion that the traffic condition of the system is improved with the packet generation rate increases. We think that our theoretical method can also be used in other transportation models with similar mechanisms, and the corresponding results can help us understand the properties of the transportation process in real-world communication system.

\section*{Method} \label{method}

To solve this model, we consider a single network first, which could give the properties of the two networks when there is no coupling. For a network with size $N$, there are $N(N-1)$ origin-destination pairs in the network. Therefore, at each time step we can find a new packet in each origin-destination pair with probability $r/(N-1)$. As we know, the betweenness of a link is the number of the shortest paths passing through it. Hence, in the free-flow state, the probability that there are $n$ new packets waiting at node $i$ for being delivered through link $i\rightarrow j$ at each time step is
\begin{equation}
p_n(b_{ij})=\binom{b_{ij}}{n} \left(\frac{r}{N-1}\right)^n\left(1-\frac{r}{N-1}\right)^{b_{ij}-n}. \label{pn}
\end{equation}
Here, $b_{ij}$ is the betweenness of link $i\rightarrow j$. When there is more than one packet in these paths, some collision may occur in the transportation of packets before they reach node $i$. As a result, $p_n(b_{ij})$ will be slightly larger than the simulation results for large $r$ and $n$. But it does not affect the theoretical results much, since $p_n$ tends to zero quickly with the increasing of $n$.

Excluding the new arriving packets, there may be some other packets waiting for being delivered through link $i\rightarrow j$ at node $i$, which remain from the last time step. We denote the number distribution of these packets as $\lambda_n(b_{ij})$. Using the two distributions $p_n(b_{ij})$ and $\lambda_n(b_{ij})$, the bandwidth usage of link $i\rightarrow j$ can be expressed as
\begin{eqnarray}
\mu_{ij}(b_{ij}) &=& \sum_{n=1}p_n(b_{ij})+p_0(b_{ij})\sum_{n=1}\lambda_n(b_{ij}) \nonumber \\
&=& 1-p_0(b_{ij})+p_0(b_{ij})[1-\lambda_0(b_{ij})] \nonumber \\
&=& 1-p_0(b_{ij})\lambda_0(b_{ij}). \label{mu}
\end{eqnarray}
Obviously, this means that when there are already some packets or some new arriving packets at node $i$ waiting for being delivered through link $i\rightarrow j$, the bandwidth of link $i\rightarrow j$ will be used at this time step.

To obtain $\mu_{ij}(b_{ij})$, we must get $\lambda_n(b_{ij})$ first, which takes the same distribution for each time step in the steady state. Thus, considering two successive time steps, it is easy to find $\lambda_n(b_{ij})$ follows the recursion formula,
\begin{eqnarray}
\lambda_0 &=& p_0\lambda_1 + (p_1 + p_0)\lambda_0, \label{l0} \\
\lambda_1 &=& p_0\lambda_2 + p_1\lambda_1  + p_2\lambda_0, \label{l1} \\
\lambda_2 &=& p_0\lambda_3 + p_1\lambda_2  + p_2\lambda_1 + p_3\lambda_0, \label{l2} \\
& \vdots & \nonumber \\
\lambda_n &=& \sum_{i=0}^{n+1}p_i\lambda_{n+1-i}, n>0. \label{ln}
\end{eqnarray}
These equations hold for any betweenness $b_{ij}$, so we omit $b_{ij}$ for simplify. In addition, as a distribution, the probability $\lambda_n$ also satisfies
\begin{equation}
\sum_{i=0}^n\lambda_{i}=1. \label{ls}
\end{equation}
The group of eqs.(\ref{l0})-(\ref{ls}) can be written as a matrix equation
\begin{equation}
P\Lambda=C, \label{ml}
\end{equation}
where $P$ is an $(n+1)\times (n+1)$ matrix,
\begin{equation}
P=\left(
\begin{array}{ccccc}
1-p_1-p_0 & -p_0  & \cdots  & 0 & 0 \\
-p_2      & 1-p_1 & \cdots  & 0 & 0 \\
\vdots    & \vdots  & \ddots & \vdots   & \vdots \\
-p_n      & -p_{n-1}  & \cdots & 1-p_1 & -p_0\\
1         & 1           & \cdots  & 1 & 1
\end{array}
\right),
\end{equation}
$\Lambda$ and $C$ are column vectors with $n+1$ entries,
\begin{eqnarray}
\Lambda&=&\left(
\begin{array}{c}
\lambda_0 \\
\lambda_1      \\
\vdots    \\
\lambda_{n-1}  \\
\lambda_n
\end{array}
\right), \\
C&=&\left(
\begin{array}{c}
0 \\
0 \\
\vdots    \\
0 \\
1
\end{array}
\right).  \label{ccc}
\end{eqnarray}
For any given $n$, we can get $\lambda_0$ from eqs.(\ref{ml})-(\ref{ccc}) by Cramer's rule. We take $n=1,2,3$ as examples,
\begin{eqnarray}
\lambda_0 &=& \frac{p_0}{1-p_1},~~n=1, \\
\lambda_0 &=& \frac{p_0^2}{(1-p_1)^2-p_0p_2},~~n=2, \\
\lambda_0 &=& \frac{p_0^3}{(1-p_1)^3-2p_0p_2(1-p_1)-p_0^2p_3},~~n=3. \label{ln3}
\end{eqnarray}
Then, substituting $\lambda_0$ into eq.(\ref{mu}), we can get the bandwidth usage. Since $p_n$ tends to zero quickly with the increasing of $n$, the bandwidth usage $\mu_{ij}$ obtained by eqs.(\ref{mu}) and (\ref{ln3}) has already agreed with the simulation results shown in Fig.\ref{f2} (b) well.

For $r\ll N$ and large $b_{ij}$, ignoring $p_n$ with $n\geq1$, it is easy to find that eq.(\ref{mu}) can be written in a simple form
\begin{equation}
\mu_{ij} = \frac{rb_{ij}}{N-1}. \label{mus}
\end{equation}
This result has also been found in ref.\cite{PhysRevE.80.066116,Ling20104825} from other perspective. Here, we want to point out that eq.(\ref{mus}) is the upper limit of the bandwidth usage, and means that all the packets do not clash before they arrive at node $i$. Therefore, the bandwidth usage obtained by eq.(\ref{mus}) will deviate from the true value for links with larger betweennesses. In addition, eq.(\ref{mus}) can also be obtained by the first moment of the distribution (\ref{pn}).

Strictly speaking, the bandwidth usage $\mu_{ij}$ obtained by eq.(\ref{mu}) will always be smaller than $1$ for $r<N-1$. To obtain the transition point indicated by the simulation, we can express the congestion condition of a link as
\begin{equation}
\mu_{ij}=1-\epsilon, \label{rc}
\end{equation}
where $\epsilon$ is a constant much smaller than $1$. This means that when the bandwidth usage $\mu_{ij}$ is close to $1$, the link will be congested. Using this equation, we can obtain the congestion condition for each link. The transition point of the system corresponds to that of the link with the largest betweenness $b_{ij}=b_{max}$. For the case shown in Fig.\ref{f2}, $b_{max}=13656$ and $N=1000$, using $\epsilon=10^{-2}$, we can find $r_c = 0.112$, which agrees with the simulation results well. In addition, we can also get the transition point $r_c = 0.073$ by eq.(\ref{mus}). Obviously, this result is much smaller than the simulation results.

For the two coupled networks, considering the two networks as two single networks, we can get the bandwidth usage of a link $\mu_{ij}$ in network $A$ and its coupled link $\mu_{ij}^{\prime}$ in network $B$ by eq.(\ref{mu}). In this way, the congestion condition of the corresponding coupled link in network $B$ is
\begin{equation}
\mu_{ij}+\mu_{ij}^{\prime}=1-\epsilon. \label{mumax}
\end{equation}
Using eq.(\ref{mu}), we have
\begin{equation}
p_0(b_{ij})\lambda_0(b_{ij}) + p_0(b_{ij}^\prime)\lambda_0(b_{ij}^\prime)=1+\epsilon, \label{rcr}
\end{equation}
Here, $b_{ij}$ is the betweenness of link $i\rightarrow j$ in network $A$ and $b_{ij}^\prime$ is the betweenness of its coupled link in network $B$. The transition point of network $B$ corresponds to $b_{ij}=b_{max}$ and $b_{ij}^\prime=\langle b\rangle$, $b_{max}$ and $b_{min}$ for random, assortative and disassortative couplings, respectively. For the networks used in the simulation of Figs.\ref{f3} and \ref{f4}, $b_{max}=13656$, $b_{min}=24$ and $\langle b\rangle=1003$. Then, we can get the transition point by letting $\epsilon=10^{-2}$, the results are $r_c^A=r_c=0.112$, and $r_c^B= 0.078$, $0.037$ and $0.109$ for random, assortative and disassortative couplings, respectively.

\bibliography{ref}

\section*{Acknowledgements}

The research of M.-B.H. was supported by the Key Research and Development Program (No.2016YFC0802508) and the National Natural Science Foundation of China (No. 11672289). The research of M.L. and B.-H.W. were supported by the National Natural Science Foundation of China (Nos. 61503355 and 11275186) and the Fundamental Research Funds for the Central Universities.

\section*{Author contributions statement}

M.L. designed the research, performed the numerical simulation, developed the theory and wrote the manuscript. M.-B.H. and B.-H.W. participated in the motivation and discussion of the results. All authors reviewed the manuscript.

\section*{Additional information}

\textbf{Competing financial interests}: The authors declare no competing financial interests.

\end{document}